\documentclass[12pt]{iopart}
\usepackage{graphicx}
\begin{document}

\title{All-fibre source of amplitude-squeezed light pulses}

\author{Markus Meissner \footnote[3]{To whom correspondence should be addressed: mmeissner@optik.uni-erlangen.de}, Christoph Marquardt, Joel Heersink, Tobias Gaber, Andr\'e Wietfeld, Gerd Leuchs, and Ulrik L. Andersen 
}

\address{Institut f\"{u}r Optik, Information und Photonik,  Max-Planck Forschungsgruppe Universit\"{a}t Erlangen-N\"{u}rnberg, Staudtstrasse 7/B2, 91058, Erlangen, Germany}

\begin{abstract}
An all-fibre source of amplitude squeezed solitons utilizing the self-phase modulation in an asymmetric Sagnac interferometer is experimentally demonstrated. The asymmetry of the interferometer is passively controlled by an integrated fibre coupler, allowing for the optimisation of the noise reduction. We have carefully studied the dependence of the amplitude noise on the asymmetry and the power launched into the Sagnac interferometer. Qualitatively, we find good agreement between the experimental results, a semi-classical theory and earlier numerical calculations\cite{schmitt98.prl}. The stability and flexibility of this all-fibre source makes it particularly well suited to applications in quantum information science.
\end{abstract}



\maketitle

\section{Introduction}
The production of squeezed states of light has attracted a wide audience within many different areas of physics. Recently, however, the arena of squeezed state applications has been dominated by the field of quantum information science. In this flourishing field, nonclassical states (squeezed as well as entangled states) are essential resources for many quantum information processing tasks such as teleportation\cite{furusawa98.sci}, cryptography\cite{ralph00.pra}, quantum computing\cite{divincenzo95.sci} and error correction\cite{braunstein98.prl}. Since these resources provide the first step in the implementation of the mentioned quantum information operations, their availability, stability and integrability is of relevance. In addition to the need for efficient generation of squeezed states, efficient state transmission between nodes of quantum operations in a quantum network is vital. The ubiquitous standard optical fibres serve as near lossless transmission lines over short distances between nodes of operations, making the infrastructure of the network efficient and simple. These same fibres can also provide the neccesary non-linearity to generate non-classical states. Therefore, an all-fibre approach to quantum information networks is promising. In this work we present a highly stable, compact and easily implementable all-fibre source of squeezed light.       

The first experimental demonstration of squeezed light, conducted in 1985 by Slusher and coworkers, was based on the process of four-wave mixing in a sodium atom beam\cite{slusher85.prl}. Shortly after this landmark experiment the first squeezing experiment exploiting the Kerr non-linearity in fibres induced by a continuous wave laser beam was performed\cite{shelby86.prl}. This experiment produced only modest squeezing due to the presence of Guided Acoustic Wave Brillouin Scattering (GAWBS)\cite{shelby85.prl}. Using ultra short laser pulses this limiting effect was reduced and experiments employing solitons\cite{rosenbluh91.prl,margalit98.ox} as well as zero-dispersion pulses\cite{bergman93.ol,bergman91.ol,bergman94.ol} were performed with greater success.          

Since the photon number is a constant of motion in a Kerr medium the number distribution of a coherent input state will remain constant during propagation. In contrast the state's area in phase space will be squeezed in dependence on the non-linearity, the injected beam power and the length of the fibre. Amplitude squeezing can not be observed unless action is taken to rotate the squeezing ellipse in phase space. Towards this end, different techniques have been devised, e.g. reflecting the quadrature squeezed field of a phase-shifting cavity to impose a phase shift on the carrier frequency, relative to the spectral RF component. The squeezed quadrature is then mapped onto the amplitude\cite{levenson85.ol}. Alternatively, Kerr effect squeezing may be turned into amplitude squeezing by spectral filtering\cite{friberg96.prl,spaelter98.oe}. Another technique was proposed by Shirasaki and Haus in 1990\cite{shirasaki90.josab} and triggered a series of new squeezing experiments based on pulsed lasers with the first implementations performed by Rosenbluh and Shelby\cite{rosenbluh91.prl} in the soliton regime, and by Bergman and Haus\cite{bergman91.ol} in the zero-dispersion regime. In these experiments pulses of equal intensity were counter propagated in a balanced Sagnac interferometer. Subsequently the two pulses destructively interfered on a 50/50 beam splitter to produce a vacuum squeezed state which was detected using a mode matched local oscillator. Since the two pulses propagated through the same fibre the system was nicely self-stabilised. Recently, an all-fibre vacuum squeezing experiment based on the same idea was sucessfully implemented and an impressive 6.1dB of squeezing was observed\cite{yu01.ol}. A linear configuration, rather than a ring geometry, which improves the mode matching to a local oscillator, has also been devised\cite{doerr93.qels}.

In contrast, the asymmetric fibre Sagnac interferometer efficiently generates directly observable amplitude squeezed states\cite{kitagawa86.pra}. Here the beam splitter of the Sagnac interferometer has unequal splitting ratio, thereby dividing the input pulse into a strong pulse that experiences the fibre non-linearity and a weak auxillary pulse which propagates through the fibre without significantly changing its statistical nature. After counter-propagating through the fibre the two pulses interfere on the beam splitter. This recombination displaces the bright pulse in phase space and by carefully choosing the input power and beam splitting ratio, directly detectable amplitude squeezing is produced. This scheme was first demonstrated by Schmitt et al.\cite{schmitt98.prl} and Krylov and Bergman\cite{krylov98.ol}. Later, the same principles were employed in a Mach-Zehnder interferometer, which allows for more degrees of freedom and thus leads to finer adjustment of the squeezing source\cite{fiorentino01.pra}. However, this configuration is not self-stabilised and an active electronic feedback loop is needed to obtain stable squeezing. 


An all-fibre integrated squeezing source is highly desirable for many modern applications using squeezed light. We have implemented an all-fibre asymmetric Sagnac interferometer with variable coupling ratio. The addition of the extra degree of freedom allows a systematic study of the dependence of the noise properties of the beam splitting ratio and enables a qualitative comparison with the semi-classical theory for a Kerr non-linear interferometer.

\section{Experimental Setup}

\begin{figure}[htbp]
	\centering
		\includegraphics[scale=0.50]{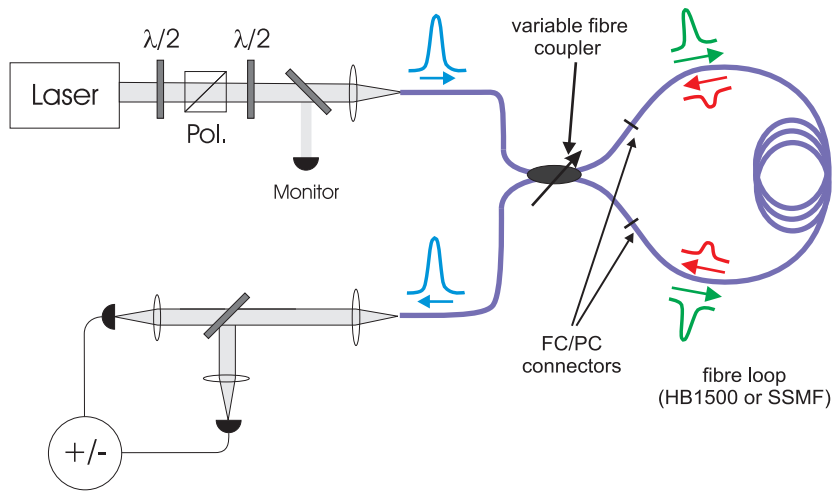}
	\caption{Experimental setup of the Sagnac fibre interferometer.}
	\label{Setup}
\end{figure}

The experimental setup is depicted in Fig.~\ref{Setup}. We used a Ti:Sapphire laser to pump an OPO to generate 100~fs pulses at 1530~nm with a repetition rate of 82~MHz. The input power to the asymmetric Sagnac loop was controlled using a half-wave plate and polarising beam splitter combination. A glass plate was used to monitor this power. A further half-wave plate rotated the beam onto one of the polarisation axes of the fibre used in the Sagnac loop. The Sagnac interferometer consisted of polarisation maintaining HB1500 fibre and a variable fibre coupler. Measurements were done with two different loop lengths: 4.8~m and 9.5~m, to check for any length dependence of squeezing. The HB1500 has a dispersion of $\beta_2=$-13.7~$ps^2/km$ and a non-linear coefficient $\gamma$ of 2.9~$(W \cdot km)^{-1}$, resulting in a soliton energy of 150 pJ. A variable fibre coupler (using HB1500 fibre from Fibercore Ltd.) was used to split the pulses into a strong and a weak pulse. Thus we were able to vary the splitting ratio from 100:0 to 50:50. The fibre lengths were connected to the coupler via FC/PC fibre connectors, introducing a linear loss of 14.4\% (0.34~dB per connector). The noise at the interferometer output was characterised using a balanced homodyne detection scheme, the sum and difference signals of which were investigated on two spectrum analysers at 20.5~MHz. These provided simultaneous measurements of the output AC noise power (sum signal) as well as the corresponding shot noise (difference signal). The measurements using this setup were carried out by recording both noise powers whilst varying the input power for a given fixed splitting ratio. This was repeated for a variety of splitting ratios between 100:0 and 70:30.\\

In a second setup we also investigated an interferometer using 50~m standard single mode fibre (Siecor SSMF) and a fused 90:10 coupler on a passively mode-locked Cr$^{4+}$:YAG laser. Apart from a fixed ratio coupler and polarisation controllers in the loop the setup was identical to that depicted in Fig.~\ref{Setup}. In this case the laser was operated at 1500~nm and produced 130~fs pulses. The SSMF has a low non-linearity, $\gamma$ is 1.05~$(W \cdot km)^{-1}$ and a dispersion of $\beta_2=$-20.3~$ps^2/km$, resulting in an energy for the first order soliton of 560 pJ. The experiments were carried out with an input pulse energy of 156 pJ. Due to the losses of the connectors and polarisation controllers the energy of the propagating pulses is even lower and below the soliton threshold of N$<$0.5. Therefore, both pulses propagating in the interferometer were spread in time due to dispersion.

\section{Experimental Results}\label{results}

\begin{figure}[htbp]
	\centering
		\includegraphics[scale=0.99]{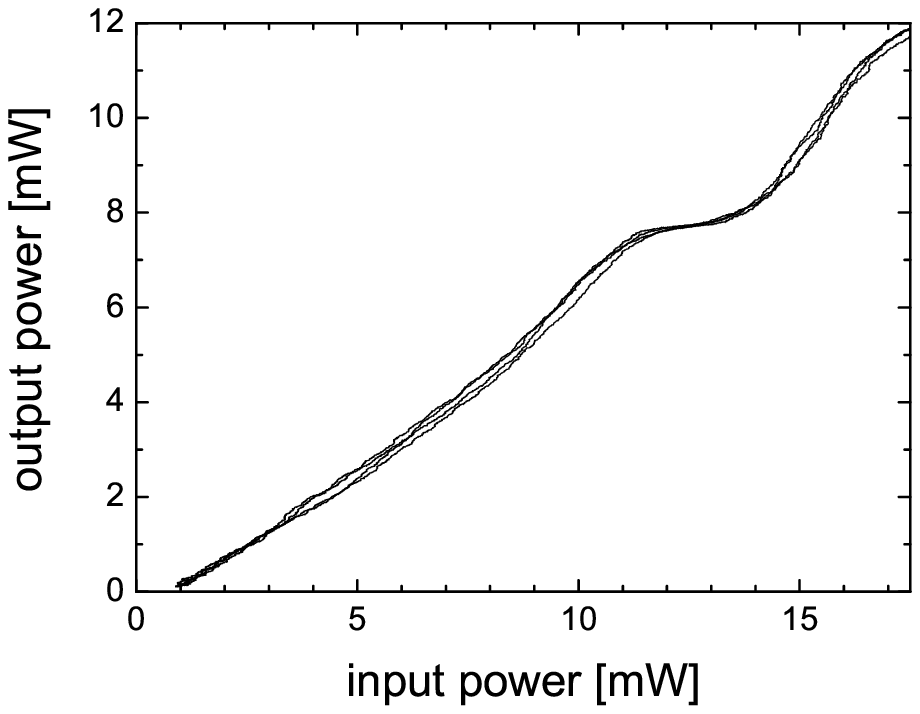}
	\caption{Non-linear input-output power transfer characteristic of the Sagnac interferometer. This data was taken using 9.5~m of HB1500 fibre for the fibre loop and a splitting ratio of 93:7.}
	\label{Kennlinie}
	\centering
		\includegraphics[scale=0.99]{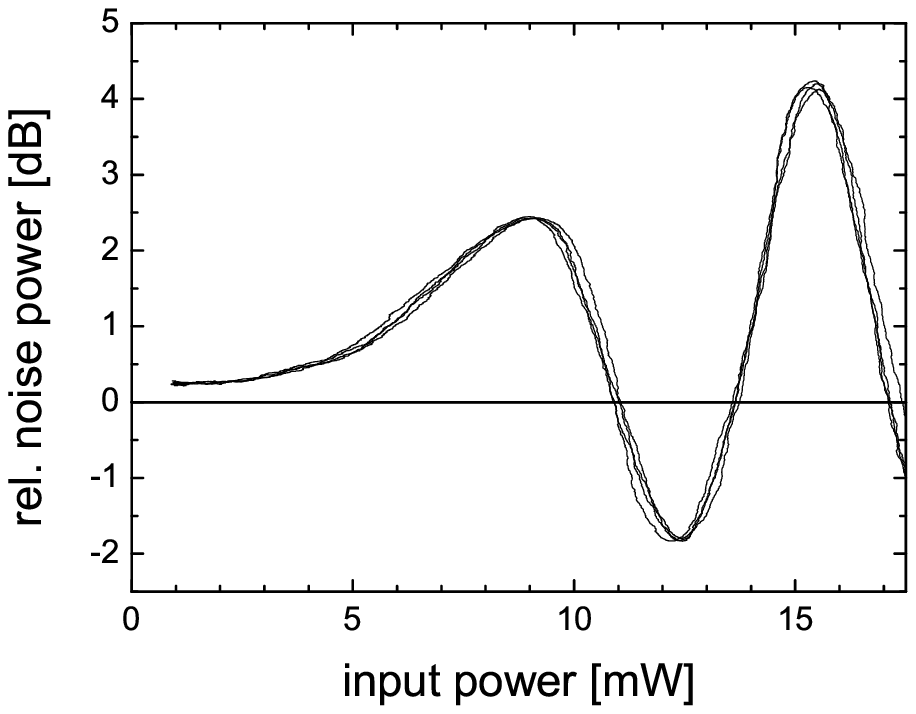}
	\caption{Noise power at the interferometer output normalized to the shot noise. Squeezing of 1.8~dB is achieved. This data was taken using 9.5~m of HB1500 fibre for the fibre loop and a splitting ratio of 93:7.}
	\label{squeezing}
\end{figure}

\begin{figure}[htbp]
	\centering
		\includegraphics{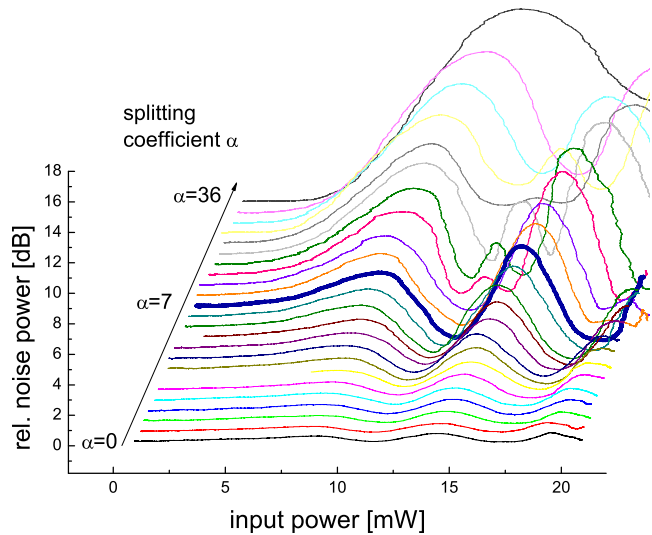}
	\caption{Waterfall diagram of the relative AC noise power at the interferometer output for different splitting ratios (note the splitting coefficient's non-linear scale), using 9.5~m of HB1500 fibre for the fibre loop. Note that both the first and second plateaus lead to squeezing.}
	\label{wasserfalldiagramm}
\end{figure}

\begin{figure}[htbp]
	\centering
		\includegraphics[scale=0.99]{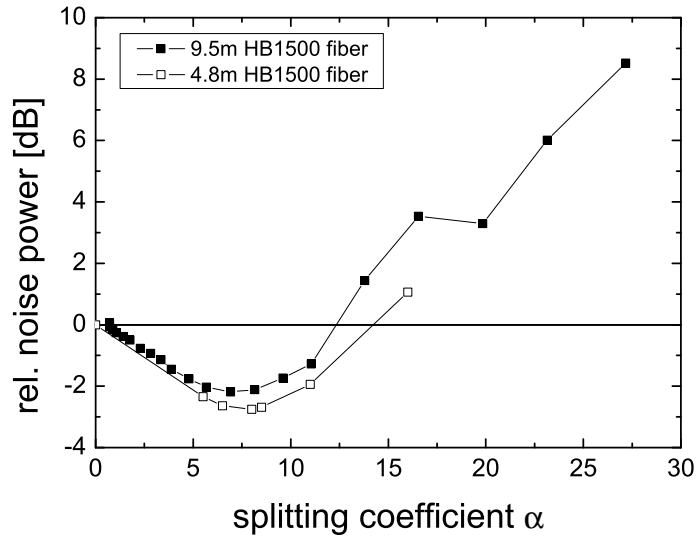}
	\caption{Relative noise power at the interferometer output in the middle of the plateau dependent on the splitting ratio. Shown are measurements taken with 4.8~m and 9.5~m of HB1500 fibre.}
	\label{squeezing_alpha_Messung}
\end{figure}

The non-linear input-output power transfer characteristic of the interferometer consisting of 9.5~m HB1500 fibre and a splitting ratio of 93:7 is shown in Fig. \ref{Kennlinie}. For input powers between 11~mW and 13.5~mW the plot exhibits a distinct plateau. A mean input power of 12 mW corresponds to a pulse energy of 146 pJ, which is very close to the energy of the first order soltion. The corresponding relative noise power to this power transfer characteristic is depicted in Fig.~\ref{squeezing}, where multiple measurement runs are shown. These results are uncorrected for linear losses or electronic dark noise. For input powers in the plateau region (11-13.5~mW), the AC noise at the output decreases below the shot noise, i.e. squeezing is observed. At the first plateau we measure $-1.8 \pm 0.2$~dB of squeezing, i.e. noise reduction. At the second plateau we find a squeezing value of $-2.0 \pm 0.2$~dB. Varying the splitting ratio changes the maximum squeezing. This is seen in the waterfall diagram of Fig.~\ref{wasserfalldiagramm}. Here the splitting coefficient is used: a splitting coefficient of 7 equals a splitting ratio of 93:7. Increasing the splitting coefficient from 0 to 7, a single squeezing dip appears on the first plateau at 12~mW. The squeezing magnitude increases with the splitting ratio, until a maximum is reached at 93:7. Further increasing the splitting coefficient reduces the squeezing, and the minimum begins to show a double dip characteristic, a result of the classical interference of the two pulses. The height of the intervening maximum tends to increase with the splitting coefficient, and even exceeds the shot noise for splitting coefficients larger than 11, i.e. for splitting ratios more symmetric than 89:11. Plotting the relative noise value that corresponds to the middle of the plateau in the transfer characteristic against splitting ratio gives rise to Fig.~\ref{squeezing_alpha_Messung}. 

We made the same measurements with a shorter fibre loop of 4.8~m, inserting pulses of 312 pJ  in the plateau region, the results of which are shown in Fig.~\ref{squeezing_alpha_Messung}. Here the maximum squeezing was $-2.4 \pm 0.2$~dB, 0.6~dB more than with the longer loop. Nevertheless, the splitting ratio dependence is very similar to the previous results. In both cases the maximal squeezing was found for a splitting ratio of nearly 93:7, as predicted in earlier numerical simulations \cite{schmitt98.prl}. The relative noise at the middle of the plateau exceeds the shot noise for splitting ratios of 89:11 for the 9.5~m loop and for 86:14 in the 4.8~m loop. It would be interesting to investigate even shorter fibres. However in this case one needs higher input powers to reach the plateau, which could not be handled by our detection system. 

\begin{figure}
	\centering
		\includegraphics[scale=0.99]{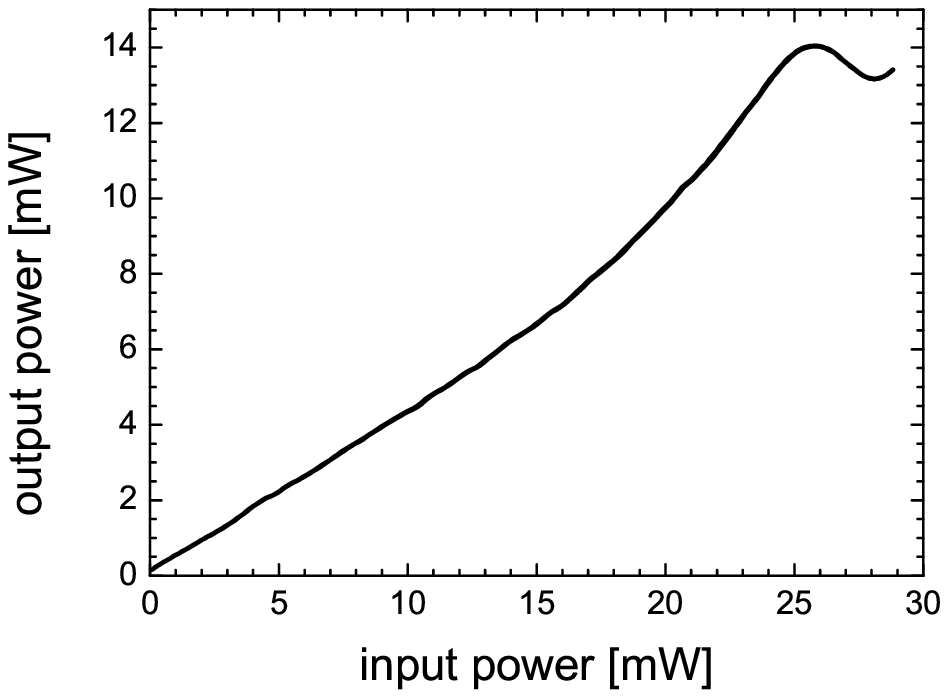}
	\caption{Non-linear input-output power transfer characteristic taken with an interferometer consisting of 50~m SSMF and a fused 90:10 coupler.}
	\label{50mSMF_kennlinie}
\centering
		\includegraphics[scale=0.99]{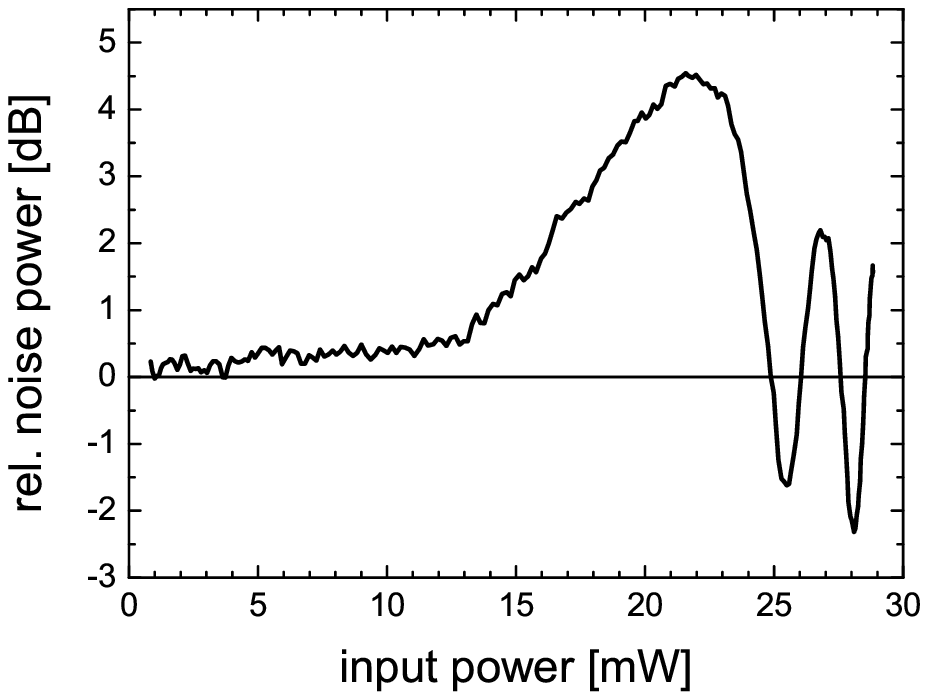}
	\caption{Relative AC noise power at the interferometer output corresponding to the transfer characteristic of Fig. \ref{50mSMF_kennlinie} (taken with an interferometer consisting of 50~m SSMF and a fused 90:10 coupler).}
	\label{50mSMF_squeezing}
\end{figure}

The non-linear input-output characteristic of the interferometer using the 50~m of SSMF is presented in Fig. \ref{50mSMF_kennlinie}. The plateau exhibits a region of negative slope, significantly larger than for the HB 1500 fibre at the same splitting ratio. Consequently, two pronounced squeezing minima occur. Based on the nonlinearity and the dispersion we estimated the soliton pulse energy for the SSMF fibre to be 560 pJ for a pulse duration of 130 fs. This is over three times larger than the soliton energy in the HB1500 fibre for the same duration, and may well explain the difference. For the SSMF fibre setup, both pulses were equally smeared out by dispersion. Thus they have a good time overlap, showing strong destructive interference. This leads to the negative slope on the plateau, which is steeper compared to the characteristics of the HB1500 and thus to two squeezing minima. These two squeezing minima are found in Fig.~\ref{50mSMF_squeezing}: $-1.6 \pm 0.2$~dB corresponding to the maximal loop power transmission and $-2.3 \pm 0.2$~dB corresponding to the minimum of the transfer characteristic.
In contrast to this, in the case of the HB1500 fibre the intense pulse is near the soliton energy of about 150 pJ, whereas the weak pulse gets smeared out by dispersion. Thus the time overlap cannot be as good as with the SSMF fibre, resulting in a weaker destructive interference and a plateau in the input-output characteristic.

\section{Discussion} 

The results presented in this paper are surprisingly well described by a simple model based on Ref.~\cite{sizmann_buch}, which provides an intuitive picture of the effects of the Kerr non-linearity on single-mode light fields. This model is visualised in phase space where $\hat{X}$ and $\hat{Y}$ are the in-phase and out-of-phase quadratures of the electromagnetic field. In phase space a coherent state can be represented by a circular uncertainty area ($\langle\Delta^{2}{\hat{X}}\rangle \langle\Delta^{2}{\hat{Y}}\rangle = 1$), displaced from the origin by a classical amplitude. The Kerr non-linearity is third order in amplitude and induces an intensity dependent phase shift in phase space. This intensity-phase interaction (self-phase modulation) transforms a coherent state into a squeezed state. This state exhibits noise reduction in some quadratures below the bound given by a coherent state. The conjugate quadratures must then be correspondingly larger than the coherent limit to fulfill the uncertainty relation. Such Kerr squeezed states have an unchanged classical amplitude, but the circular area of uncertainty has been "squeezed " into a crescent one\cite{kitagawa86.pra}. In our case these crescent uncertainty regions can be treated as ellipses as the classical amplitude ($n \approx 10^9$ photons per pulse) is much larger than the fluctuations ($\Delta^{2}n \approx \sqrt{n}$). This squeezing, however, cannot be measured by direct detection, and thus requires a reorientation of the ellipse in phase space before becoming measurable in the amplitude quadrature. This can be accomplished by interference, described here by vector addition.

\begin{figure}[htbp]
	\centering
		\includegraphics[scale=0.50]{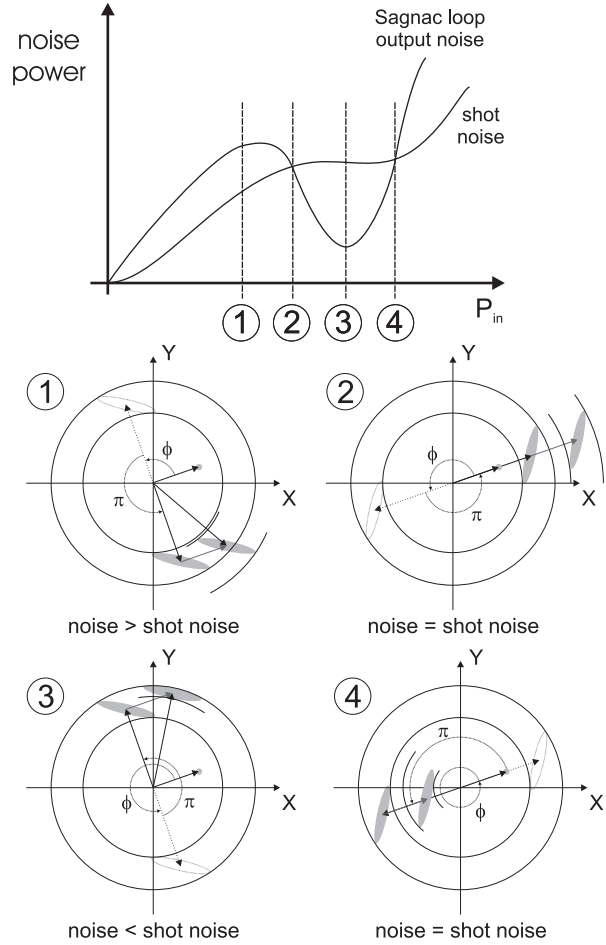}
	\caption{Schematic explanation of the noise power at the Sagnac loop output in dependence on the loop input power (X and Y denote the unnormalised quadratures of the electromagnetic field). Here $\phi$ is the relative phase between the weak and strong pulses; the constant phase shift of $\pi$ of the strong pulse on the beam splitter is also shown. (1) $\phi<\pi$ exhibits output noise above the shot noise, (2) $\phi=\pi$ leads to shot noise at the fibre loop output, (3) $\phi\approx1.5\pi$ shows maximal squeezing, (4) $\phi=2\pi$ leads to shot noise at the output.}
	\label{Pin}
\end{figure}

The dependence of the relative noise power on the input power of the Sagnac interferometer, as depicted in the phase space diagrams in Fig.~\ref{Pin}, can be explained as follows: Increasing the power for a fixed splitting ratio from zero, we find for low powers a small non-linear phase shift between the strong and the weak pulse, $\phi$, leading to a partially constructive interference with the weak pulse, bearing the $\pi$ phase shift of the bright beam on the beam splitter in mind. Here the noise level after the Sagnac loop lies above the shot noise, depicted in Fig.~\ref{Pin} subfigure (1), since the projection of the ellipse onto the amplitude quadrature is greater than that of the corresponding coherent state. Further increasing the input power to the fibre loop increases $\phi$ due to the Kerr non-linearity.  When this phase shift becomes $\pi$ complete constructive interference occurs, seen in Fig.~\ref{Pin} part (2). Thus the ellipse is simply shifted to a larger amplitude, and the output pulse noise characteristics remain equal to the shot noise. In the case of $\phi=2\pi$ destructive interference is observed, but again the noise characteristics are unchanged, being equal to the shot noise, as depicted in (4). For one specific input power, however, squeezing can be obtained. If the relative nonlinear phase shift is $\phi\approx1.5\pi$, the ellipse is shifted such that the minor axis of the ellipse lies parallel to the amplitude of the pulse (as shown in subfigure (3)). This specific input power coincides with the middle of the plateau of the non-linear input-output power transfer characteristic of the fibre loop.

\begin{figure}[htbp]
	\centering
		\includegraphics[scale=1.0]{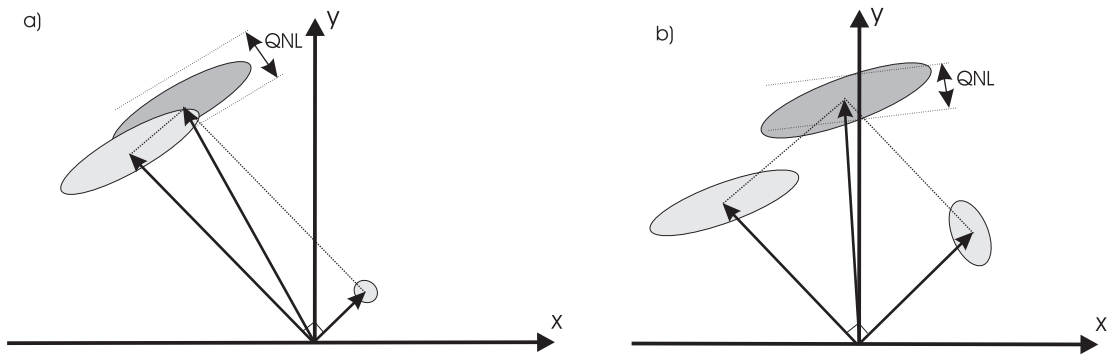}
	\caption{Schematic explanation of the splitting ratio dependence of the noise power at the Sagnac interferometer output: a) For a beam splitting ratio with a large asymmetry the Kerr-squeezed state is reorientated to yield amplitude squeezing, identical to the visualisation in Fig. 8 part 3. b) The beam splitter is more symmetric leading to an inapropriate displacement of the Kerr-squeezed beam. In this case the amplitude noise of the resultant field is above the quantum noise limit. x and y denote the unnormalised quadratures of the electromagnetic field}
	\label{alpha}
\end{figure}

To qualitatively understand the mechanism at work in the asymmetric Sagnac loop in more details, we consider the pictorial phase space description in Fig.~\ref{alpha}. The initial Kerr squeezed beam, without any interference, exhibits a noise power equal to the shot noise in its amplitude. Interfering this squeezed beam with a weak, auxillary beam, oscillating 90° out of phase, shifts the uncertainty ellipse.  For a particular power, or in the case of a Sagnac loop for a particular splitting ratio, the ellipse orientation is such that maximum squeezing is seen in the amplitude quadrature, Fig.~\ref{alpha}a.  Further increasing the auxillary power, or splitting ratio, decreases this squeezing until it begins to rise above the shot noise, i.e. the anti-squeezing becomes visible again, seen in Fig.~\ref{alpha}b.

\begin{figure}[htbp]
	\centering
		\includegraphics[scale=0.99]{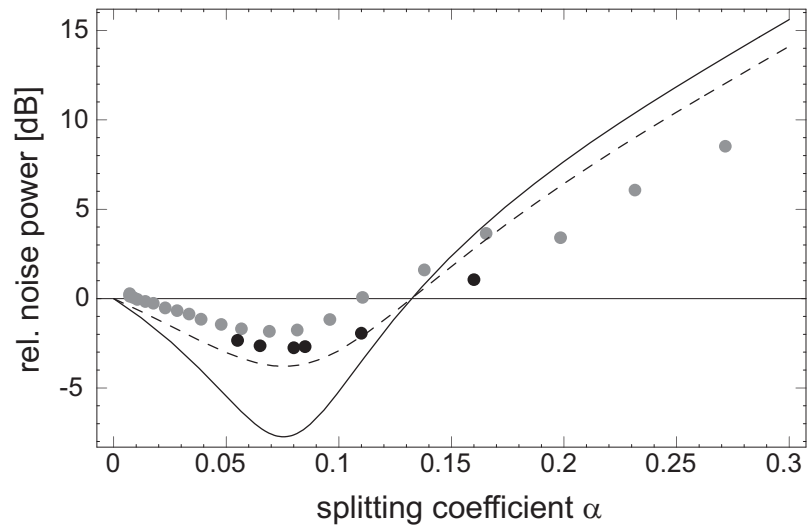}
	\caption{Theoretically predicted relative noise power at the interferometer output as a function of the beam splitting ratio.}
	\label{rechnung}
\end{figure}

Based on the single mode pictorial squeezing model of Fig.~\ref{alpha}, rough estimates were calculated to verify the experimental results. To simplify the derivation we consider only the middle of the plateau of the input-output power plot of the Sagnac loop in Fig.~\ref{Kennlinie} corresponding to a relative phase shift between the two beams of ~1.5$\pi$. The acquired non-linear phase shifts are then:
\begin{equation}
\phi_{NL1}=(1-\eta)\frac{3\pi}{2(2\eta-1)}, \phi_{NL2}=\eta\frac{3\pi}{2(2\eta-1)}
\end{equation}
for the strong and weak beam respectively. $\eta$ denotes the reflectivity of the beam splitter of ratio $(1-\eta):\eta$. As a result of the non-linear phase shifts, both beams become quadrature squeezed and can be described by the variances\cite{sizmann_buch}:
\begin{equation}
V_i^\theta=\sin^2\theta((\cot\theta+2\phi_{NLi})^2+1)
\end{equation}
where $i=1,2$ and $\theta$ is the quadrature angle. $\theta=0$ corresponds to the variance of the amplitude of the individual beams and for this case we easily find $V_i^0=1$ which is the quantum noise limit as expected from a Kerr-squeezed beam. The Kerr-squeezed ellipses are projected onto the radial direction, or the amplitude quadrature, of the resultant field with the angles $\theta=\pm \arctan(\eta/(1-\eta))$ respectively, and the relative noise power is determined by adding the two noise contributions incoherently, weighted by the beam splitting ratio. Using this model calculations of the maximum squeezing for splitting ratios in the range from 100:0 to 70:30 were performed, shown in Fig.~\ref{rechnung}. The dashed line in Fig.~\ref{rechnung} shows the calculated squeezing degraded by the 30\% loss introduced in the whole setup. The shape of this plot agrees qualitatively with the uncorrected experimental data presented in the same figure. The model predicts optimum squeezing for a beam splitting ratio of 92.5:7.5, which is in agreement with the measured ratio -- 93:7. According to the model the noise power exceeds the quantum noise limit for ratios above 88:12. Experimentally we measured this intersection at the ratios of 89:11 (9.5~m HB1500) and 86:14 (4.8~m HB1500), very similar to the theoretical value. Quantitative agreement between the squeezing values of earlier calculations and the simplistic model presented here with the experimentally results is however lacking. Earlier full scale quantum calculations predicted -11.0~dB \cite{wernerprl.98}; the simplistic model presented here predicts -7.7~dB of squeezing, both of which differ from the maximum measured squeezing of approximately -4~dB (measured value of -2.4~dB corrected for 30\% loss).

Our simplistic single mode model has the following weaknesses:
a) assuming the uncertainty region to be of elliptical form, 
b) using a linearized non-linear phase shift,
c) assuming a constant relative phase of 1.5$\pi$ as the angle for optimal squeezing.
d) neglecting various non-linear effects (apart from self-phase-modulation) have been,
e) ignoring Guided Acoustic Wave Brillouin Scattering (GAWBS), and
f) disregarding imperfect interference at the the beam splitter.
Because of these assumptions we stress that the simplistic model presented here yields only a rough estimate of the noise power of the output field. For a more complete description of the experiment one must resort to the more rigorous numerical quantum calculations presented in \cite{wernerprl.98,schmitt98.prl}. These calculations used the quantum nonlinear Schr\"{o}dinger equation, implementing only the lowest order terms. The symmetrised split-step Fourier method was employed to propagate the pulses through the Sagnac interferometer. The inclusion of the Raman effect has been shown to have little impact on the squeezing formation\cite{corney}.

A possible mechanism for the quantitative discrepancy between the theoretical predictions and the experimetal results are linear losses in the fibre and setup. These were measured to be approximately 30\% (taking into account the quantum efficiency of the photodiodes of about 90 \%). The experimental values of Fig.~\ref{squeezing_alpha_Messung} and the calculated degraded squeezing (dashed line) can be compared in Fig~\ref{rechnung}. Another possibility is a poor interference in the variable coupler, since other amplitude squeezing experiments (i.e.\cite{schmitt98.prl}) using the Sagnac interferometer in free-space configuration have seen noticeably more squeezing\cite{schmitt98.prl,SchmittJaBe}. Further, the results of Fig.~\ref{50mSMF_squeezing} using a fused fibre coupler also exhibited weak squeezing. It is however unlikely, that the interference quality is solely responsible, as the interference effects of the Sagnac loop in the output power are as apparent as in other experiments using this device. A final possibility is that Guided Acoustic Wave Brillouin Scattering (GAWBS) decreased the squeezing by coupling acoustical vibration modes of the fibre into the amplitude quadrature. The exact cause of the deviation between theory and experiment cannot be named at this point. Nevertheless, the qualitative agreement between both the earlier rigorous calculations, the simplistic pictorial model and the experimental results is good, all giving an optimum splitting ratio of approximately 93:7.

\section{Conclusion}

We have presented an experimental as well as a theoretical analysis of an all-fibre non-linear Sagnac interferometer that produces amplitude squeezed states of pulsed light. An integrated variable fibre-coupler allowed for a full characterisation of the squeezing dependence on the beam splitting ratio, and the optimum squeezing was found to be at a splitting ratio of approximately 93:7. The position of this squeezing optimum was surprisingly well described by a simplistic single mode model. Although this model yielded qualitative agreement with the experimental data of our
squeezing setup and earlier numerical calculations, it was unable to provide an adequate quantitative description. An improved quantitative description of the squeezing process inside a fibre must account for several additional linear and non-linear effects. We believe that the presented all-fibre source of squeezed pulses will be a key element in future quantum information networks.

\ack{This work has been supported by the Schwerpunkt programm 1078 of the Deutsche Forschungsgemeinschaft and the network of competence QIP of the state of Bavaria (A8). ULA also acknowledges funding from the Alexander von Humboldt Foundation. We gratefully acknowledge useful discussions with Oliver Gl\"ockl and Stefan Lorenz and the help of A\'ska Doli\'nska with the figures.}


\begin{thebibliography}{}
\bibitem{furusawa98.sci} Furusawa A, S\o rensen J L, Fuchs C A, Kimble H J, Polzik E S, Science {\bf 282}, 706 (1998)
\bibitem{ralph00.pra} Ralph T C, Phys. Rev. A, {\bf 61}, 010303 (2000)
\bibitem{divincenzo95.sci} Di Vincenzo D P, Science {\bf 270}, 255 (1995)
\bibitem{braunstein98.prl} Braunstein S L, Phys. Rev. Lett. {\bf 80}, 4084 (1998)
\bibitem{slusher85.prl} Slusher R E, Hollberg L W, Yurke B, Mertz J C, and Valley J F, Phys. Rev. Lett. {\bf 55}, 2409 (1985).
\bibitem{shelby86.prl} Shelby R M, Levenson M D, Perlmutter S H, DeVoe R G and Walls D F, Phys. Rev. Lett. {\bf 57}, 691 (1986).
\bibitem{shelby85.prl} Shelby R M, Levenson M D, Bayer P W, Phys. Rev. Lett. {\bf 54}, 939 (1985) 
\bibitem{rosenbluh91.prl} Rosenbluh M, Shelby R M, Phys. Rev. Lett. {\bf 66}, 153 (1991)
\bibitem{friberg96.prl} Friberg SR, Machida S, Werner MJ, Levanon A and Mukai T, Phys. Rev. Lett. {\bf 77}, 3775 (1996).
\bibitem{spaelter98.oe} Sp\"{a}lter S, Burk M, Str\"{o}ssner, Sizmann A and Leuchs G, Opt. Express {\bf 2}, 77 (1998).
\bibitem{margalit98.ox} Margalit M, Yu C X, Ippen, E P, Haus H A, Opt. Exp. {\bf 2}, 72 (1998)
\bibitem{bergman93.ol} Bergman K, Doerr C R, Haus H A, Shirasaki M, Opt. Lett. {\bf 18}, 643 (1993)
\bibitem{bergman91.ol} Bergman K, Haus H A, Opt. Lett. {\bf 16}, 663 (1991)
\bibitem{bergman94.ol} Bergman K, Haus H A, Ippen E P, Shirasaki M, Opt. Lett. {\bf 19}, 290 (1994)
\bibitem{levenson85.ol} Levenson M D, Shelby R M, Perlmutter S H, Opt. Lett. {\bf 10}, 514 (1985). Galatola P, Luigiato L A, Porreca M G, Tombesi P, Leuchs G, Opt. Comm. {\bf 85}, 95 (1991)
\bibitem{shirasaki90.josab} Shirasaki M, Haus H A, Jour. Opt. Soc. Am. B, {\bf 7}, 30 (1990)
\bibitem{doerr93.qels} Doerr C R, Lyubomirsky I, Lenz G, Paye J, Haus H A, Shirasaki M, QELS {\bf 12}, OSA Technical Digest Series, 282 (1993)
\bibitem{yu01.ol} Yu C X, Haus H A, Ippen E P, Opt. Lett. {\bf 26}, 669 (2001)
\bibitem{kitagawa86.pra} Kitagawa M, Yamamoto Y, Phys. Rev. A {\bf 34}, 3974 (1986)
\bibitem{schmitt98.prl} Schmitt S, Ficker J, Wolff M, K\"onig F, Sizmann A, and Leuchs G, Phys. Rev. Lett. {\bf 81}, 2446 (1998)
\bibitem{krylov98.ol} Krylov D, Bergman K, Opt. Lett. {\bf 23}, 390 (1998)
\bibitem{fiorentino01.pra} Fiorentino M, Sharping J E, Kumar K, Levandovsky D, Vasilyev M, Phys. Rev. A, {\bf 64}, 031801(R) (2001)
\bibitem{sizmann_buch} Sizmann A, Leuchs G, Progress in Optics 39, 376-486 (1999)
\bibitem{wernerprl.98}M. Werner, Phys. Rev. Lett., {\bf 81}, 4132 (1998)
\bibitem{corney}J. F. Corney and P. D. Drummond, JOSA B 18, 153-161 (2001)
\bibitem{SchmittJaBe} S. Schmitt, J. Ficker, A. Sizmann and G. Leuchs, Annual Report 1998, Lehrstuhl f\"ur Optik, Friedrich-Alexander-Universit\"at Erlangen-N\"urnberg, 63 (1999)
\end{thebibliography}
\end{document}